\documentclass[a4paper,
               ]{jacow}
\usepackage{pdfpages,multirow,ragged2e} %
\makeatletter%
	\ifboolexpr{bool{xetex}}
	 {\renewcommand{\Gin@extensions}{.pdf,%
	                    .png,.jpg,.bmp,.pict,.tif,.psd,.mac,.sga,.tga,.gif,%
	                    .eps,.ps,%
	                    }}{}
\makeatother

\ifboolexpr{bool{xetex} or bool{luatex}} 
 {}                                      
 {\usepackage[utf8]{inputenc}}           

\usepackage[USenglish]{babel}

\begin{document}

\title{Exploiting Non Redundant Aperture Interferometry as a Diagnostics Tool for Synchrotron Light Characterization }

\author{L. Torino\thanks{ltorino@cells.es}, U. Iriso, ALBA CELLS, Cerdanyola del Vall\'es, Spain \\
C.~Carilli, National Radio Astronomy Observatory (NRAO), Socorro - NM, USA \\
B.~Nikolic, U. of Cambridge, Cambridge, UK and 
N. Thyagarajan, CSIRO, Bentley WA, Australia  }
	
\maketitle

\begin{abstract}
  
We recently introduced a novel interferometric method inspired by radio astronomy, utilizing a Non-Redundant Aperture (NRA) mask with self-calibration to fully characterize the two dimensional transverse shape of electron beams from a single-shot interferogram.
This paper reports the latest advancements in this technique, in particular, the potential of this method for resolving beam halos superimposed on a well-defined Gaussian beam core. We also summarize a new data analysis approach based on closure amplitudes, which removes the need for self-calibration, and we demonstrate the method’s applicability to wavefront sensing.

\end{abstract}

\section{Introduction}

In recent publications \cite{Nikolic:2024bfx, Carilli24, Iriso:procibic2024-tup56}, we introduced a new technique to fully characterize the electron beam in a single interferogram produced by the synchrotron radiation and a Non-Redundant Aperture (NRA) mask with several holes. 
This interferogram is analyzed in Fourier space with the so-called "self-calibration" method, which 
leverages the information from numerous aperture combinations to reconstruct the electron beam shape and corrects for non-uniform mask illumination using gain fitting. Widely used in radio astronomy, this technique has been applied for the first time with visible light to characterize the ALBA electron beam.

The experimental setup is a modified form of two-aperture Synchrotron Radiation Interferometry (SRI) \cite{Mitsuhashi:1998em}. It uses a lens, filters, and a CCD camera \cite{LT:IBIC14} but replaces the typical two-aperture mask with an NRA mask, allowing for simultaneous interference measurements from multiple pairs of apertures, or "baselines" \cite{1980SPIE..231...18S, 1987Natur.328..694H}.

\begin{figure}[!h]
    \centering
    \includegraphics[width=.99\linewidth]{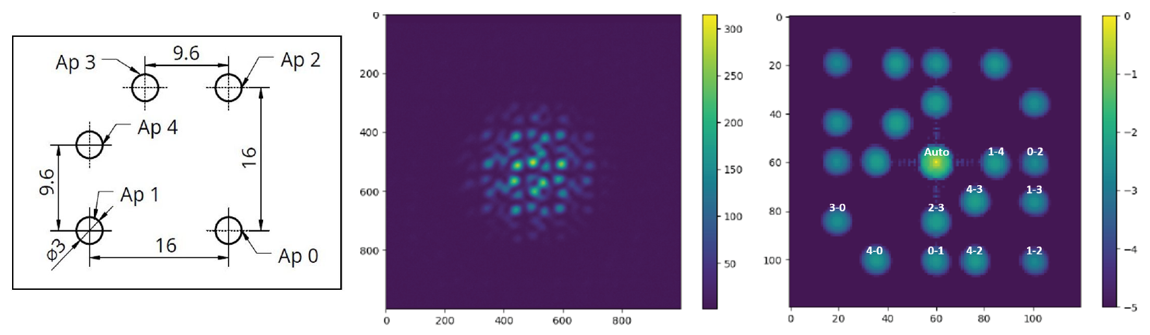}
    \caption{Example of a 5-hole NRA mask (left), its interference pattern (middle) and the 2D FFT (right).}
    \label{fig:example}
\end{figure}

Figure~\ref{fig:example} shows an example of the 5-hole mask (left), its interference pattern (middle), and the amplitude of the 2D-FFT image of the interferogram (right): 
this process reveals a series of spots, or samples, in the complex Fourier ($u,v$) plane. Note that he total number of independent samples is determined by the number of apertures ($N$) in the mask using the formula $N(N-1)/2~+~1$.

In the Fourier space, each sample provides specific information. In particular, the central sample (u,v = 0,0), represents the autocorrelation and is proportional to the total light intensity from all the holes. The other samples are generated by every possible baseline (pair of holes). A samples's location in the Fourier space corresponds to the physical distance and orientation of its generating baseline. The intensity of each sample directly corresponds to the visibility amplitude, of the interference pattern for that particular baseline. This method allows for a comprehensive reconstruction of the beam profile by measuring the complex visibilities for multiple orientations and distances at the same time.

A key advantage of this technique is its ability to account for variations in the illumination of each aperture. What we in fact measure is the visibility $V_{ij}$, that is generated by the interference between light passing the apertures $i$ and $j $, while the information we need to reconstruct the beam shape is the spatial coherence $\gamma_{ij}$. The relation between this two quantity depends on the gain $G_i$ which is the proportional to the square root of the illumination of the aperture $i$:
\begin{equation}
   \|V_{ij}\|  =  \gamma_{\rm ij} \|G_i\| \|G_j\|.
   \label{eq:vis}
\end{equation}
The electron beam is assumed to be an ellipse (or a 2D Gaussian), defined by its major and minor axes and its tilt angle. By adding the autocorrelation information
\begin{equation}
     \|V_{\rm auto}\|  =  \sum_i  \|G_i\|^2 ~,
\end{equation}
the gains $G_i$ can be treated as fitting parameters. This correction for illumination variations is necessary to obtain an accurate and complete reconstruction of the beam profile in the Fourier space:
\begin{equation}
  \gamma(u,v) = V \exp[- (a u^2 + b v^2 + 2cuv) ],
\end{equation}
where $u$ and $v$ are the coordinates of the Fourier space and $a$, $b$, $c$ are the standard parameter of a 2D Gaussian. The $\sigma_{\rm u,v}$s and the angle of the 2D Gaussian  can be obtained following \cite{wikipedia} and $\sigma$s in the real space are calculated using the following conversion:
\begin{equation}
      \sigma_{\rm x,y} =  \frac{\lambda L}{2 \pi \sigma_{\rm u,v}}~,
      \label{eq:conv}
\end{equation}
where $\lambda$ is the wavelength of the used light and $L$ is the distance between the source and the aperture.

This report shows an alternative data analysis and explore the advantages of processing images in the Fourier domain to find several applications of NRA interferometry.

\section{Closure Amplitude Data Analysis}

Alternately, a way to bypass the issue related with non-uniform illumination is to use a data analysis based on closure amplitudes: more details can be found in \cite{Thyagarajan:2025pue}. This technique was developed in radio-astronomy to overcome effects due to differing antenna gains. The idea is to use a closed loop of four aperture elements, called \textit{quads}, defined as follows:
\begin{equation}
    A_{ijkl} = \frac{\|V_{ij}\| \|V_{kl}\|}{\|V_{ik}\| \|V_{jl}\|} = \frac{\|\gamma_{ij}\| \|\gamma_{kl}\|}{\|\gamma_{ik}\| \|\gamma_{jl}\|}.
\end{equation}
Given the relation between $V$ and $\gamma$ as defined in Eq. \ref{eq:vis}, we eliminate gain dependency. The number of independent \textit{quads} is $N_q = N(N-3)/2 $, where $N$ is the number of apertures. Thus, for a 5-hole mask, which means $N_q=5$, we get 5 coherence measurements. These measurements allow us to fit a 2D Gaussian with 3 independent parameters. You can then use the conversion from Eq. \ref{eq:conv} to determine the beam size.

\begin{figure}[!h]
    \centering
    \includegraphics[width=.95\linewidth]{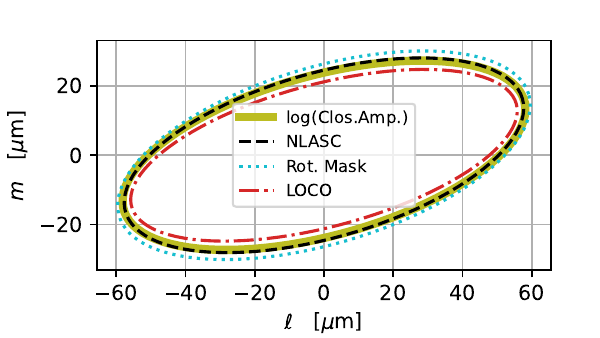}
    \caption{Electron beam measurement comparison between the NRA with closure amplitude, the NRA 5-hole mask with self calibration, the rotating 2-hole SRI, and LOCO.}
    \label{fig:clamp}
\end{figure}

Figure \ref{fig:clamp} shows the measurement of the beam size of the ALBA electron beam. Data were taken with a rotating 2-hole mask SRI setup \cite{PhysRevAccelBeams.19.122801} (light blue dashed line) and an NRA 5-hole mask analyzed using self-calibration (black dashed line), and the closure amplitude technique (yellow line). The results are comparable with each other and with respect to the theoretical expectation provided by LOCO\cite{Safranek:1997mra} (red dashed line).

\section{Waveform Sensing}

Wavefront sensing measures how light paths are deformed by optical elements or air turbulence. To measure the waveform distortion we generate an interferogram using an NRA mask. This interferogram is then analyzed with self-calibration to determine the complex gain of each hole in the mask. The complex gains include both amplitudes (i.e. illumination pattern), and phases (i.e. photon path-lengths). By comparing the gain phase ($\phi_G$) differences between each hole relative to a reference position in the mask, or reference hole, we can directly measure the variations in the path length ($\delta L$) caused by wavefront distortions:
\begin{equation}
\rm \delta L = \lambda \times (\phi_G /360^\circ). \label{eqn:path}   
\end{equation}

A full description of the technique can be found in \cite{Carilli:2025dgs}. 
To quantitatively validate the technique, we inserted a rotating mirror into the optical path at ALBA, and measured the wavefront planar solutions for a series of mirror tilts.  Data were taken with blue light ($\lambda=400\,$nm).

\begin{figure}
    \centering
    \includegraphics[width=0.8\linewidth]{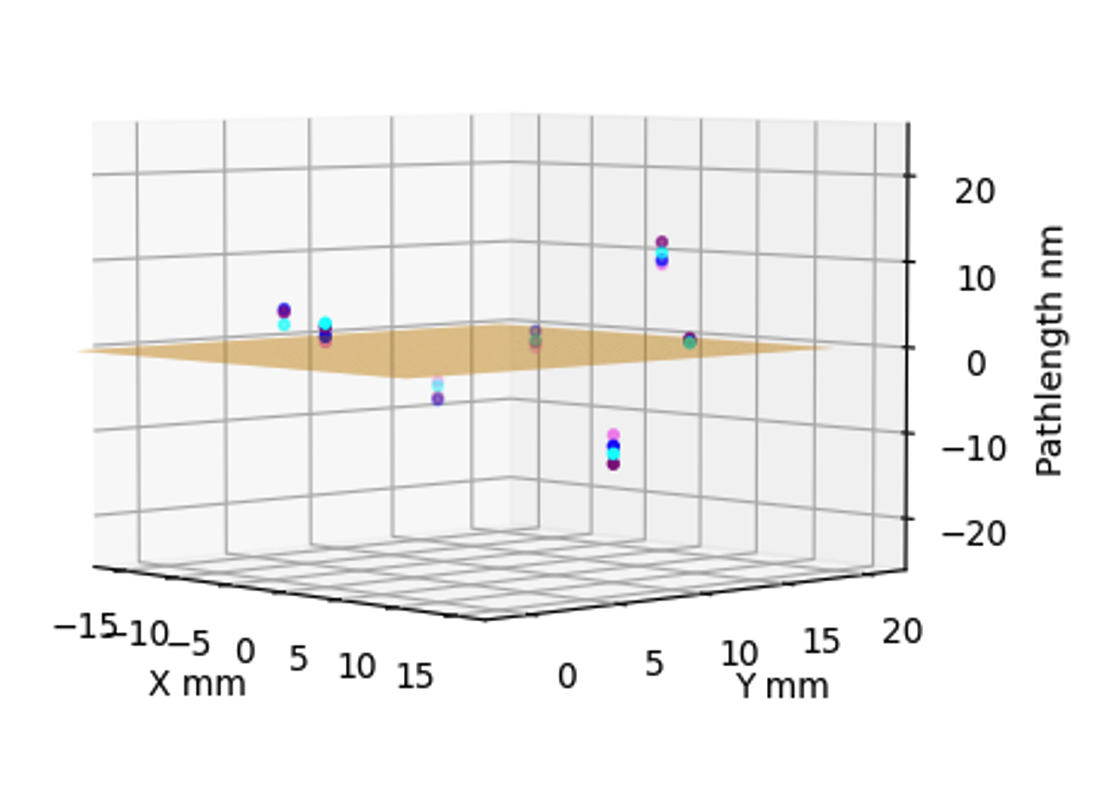}
    \includegraphics[width=0.8\linewidth]{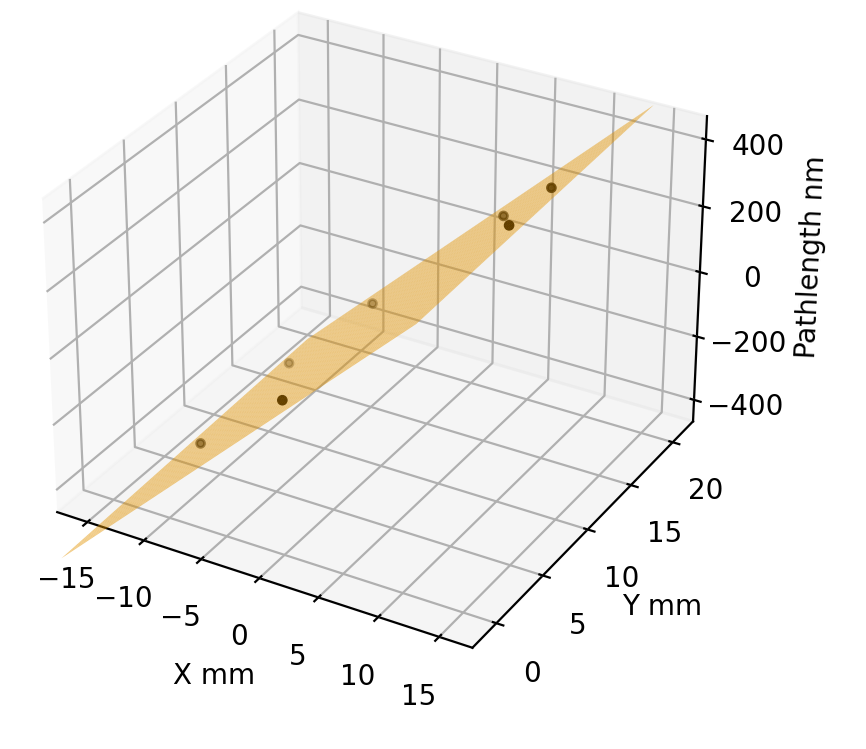}
    \caption{Top: Mean (static) residuals after subtracting the best fit plane to all the tilted mirror experiments, including no tilt. Mean residuals up to 12~nm are seen. These non-planar distortions are reproduce between experiments to $\pm 1$~nm  (top). Bottom: planar fit to the $4''$ tilt data.}
    \label{fig:TiltMir}
\end{figure}

We used three different methods to measure the tilt and compared the results. The first method involved measuring the shift of the Airy disk's center on the CCD. The second method involved looking at phase gradients across each u,v sample. The third method employed the gain phases from self-calibration, converted to photon path-length. The results from all three methods were comparable (see Table \ref{tab:restilt}). A plot of the data and planar fit is shown for one of the mirror tilts in Fig. \ref{fig:TiltMir}. The gain phase method reproduces the wavefront tilts to within $0.1''$ accuracy ($5\times 10^{-7}$ rad). We also find the static non-planar residuals for the path-lengths to each hole reproduced to within $\sim$1~nm for all the experiments (Fig.~\ref{fig:TiltMir}). 

\begin{table*}[t]
\centering
\footnotesize
\caption{Wavefront Tilt}
\begin{tabular}{lcccccc} 
  \hline
  \hline
  Mirror & Selfcal & Image Centering  & u,v sample slope & Selfcal & Image Centering  & u,v sample slope \\
rotation & X arcsec & X arcsec & X arcsec & Y arcsec & Y arcsec & Y arcsec \\
\hline
$1''$ & $2.15\pm 0.09$ & $2.30\pm 0.13$ & $2.15 \pm 0.09$ & $0.56 \pm 0.09$ & $0.66\pm 0.13$ & $0.70 \pm 0.09$ \\
$2''$ & $3.27 \pm 0.12$ & $3.21\pm 0.14$ & $3.17\pm 0.12$ & $-0.07\pm 0.13$ & $-0.01\pm 0.14$ & $0.06 \pm 0.12$ \\
$3''$ & $4.88\pm 0.1$ & $4.77\pm 0.13$ & $4.78\pm 0.12$ & $0.08\pm 0.1$ & $-0.04\pm 0.13$ & $0.11\pm 0.11$ \\
$4''$ & $6.96\pm 0.1$ & $6.92\pm 0.10$ & $6.95\pm 0.10$ & $0.24 \pm 0.1$ & $0.27\pm 0.10$ & $0.39\pm 0.10$ \\
\hline
\hline
\vspace{0.1cm}
\end{tabular}
\label{tab:restilt}
\end{table*}

\section{Double Gaussian Reconstruction}

This study investigates the use of the NRA technique to reconstruct non-Gaussian beams, a method that could be used for beam halo measurement.

To simulate a beam halo, we used the ALBA bunch-by-bunch (BBB) system \cite{Olmos:2016uwv} to create a double Gaussian beam. 
This was achieved by exciting a portion of the electron bunches at the resonant frequency ($Q_v+Q_s$), where $Q_v$ and $Q_s$ are the vertical and the synchrotron tune. 

The beam shape is then parametrized by the sum of two Gaussian beams with different beam sizes, where the amplitudes of each Gaussian, named $A_{\rm E}$ and $A_{\rm NE}$ for the excited and non-excited beam, follow the relationship: 




\begin{equation}
\frac{A_{\rm E}}{A_{\rm NE}} = \frac{\sum_{i=1}^{N_{\rm E}} Q_i}{\sum_{i=1}^{N_{\rm NE}} Q_i} ~,
\label{eq:ratio}
\end{equation}

\noindent where $N_{\rm E}$ and $N_{\rm NE}$ are the number of excited and non-excited bunches, respectively, and $Q_i$ is the charge of the bunch "i".

\begin{figure}
\centering
\centerline{\includegraphics[width=0.9\linewidth]{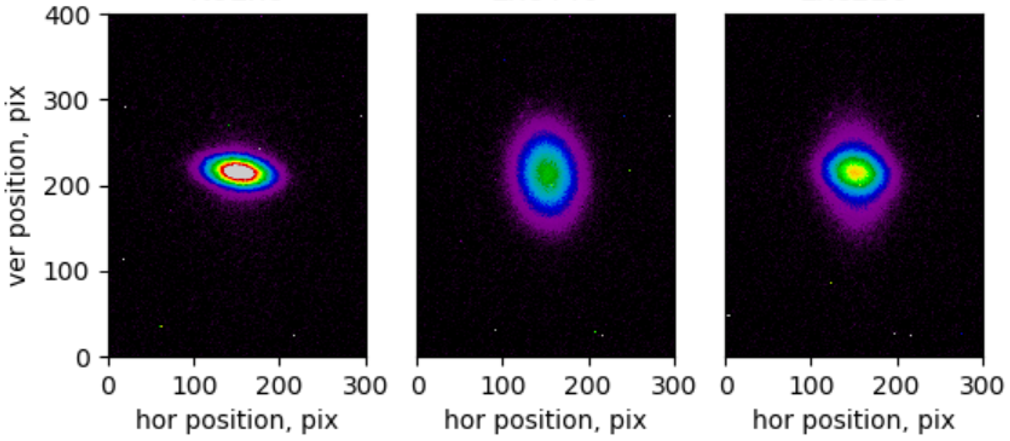}}
\caption{X-ray pinhole images of the beam in normal operation (left), compared with the beam with all bunches excited (center) and the beam with only half of the bunches excited (right).}
\label{fig:FE34}
\end{figure}

A long-exposure CCD camera (approximately milliseconds) captures the resulting combined beam. Fig. \ref{fig:FE34} shows the pinhole images at FE34 for the original, fully excited, and half-excited beams \cite{UI:IBIC22}.
We then measure the beam sizes of both the main Gaussian and the halo Gaussian, as well as the intensity ratio of the two beams. This ratio should correspond to the ratio of charges in Eq. \ref{eq:ratio}, which we use as a figure of merit to estimate our ability to measure the two components of the beam.

\begin{figure}
\centering
\centerline{\includegraphics[width=0.9\linewidth]{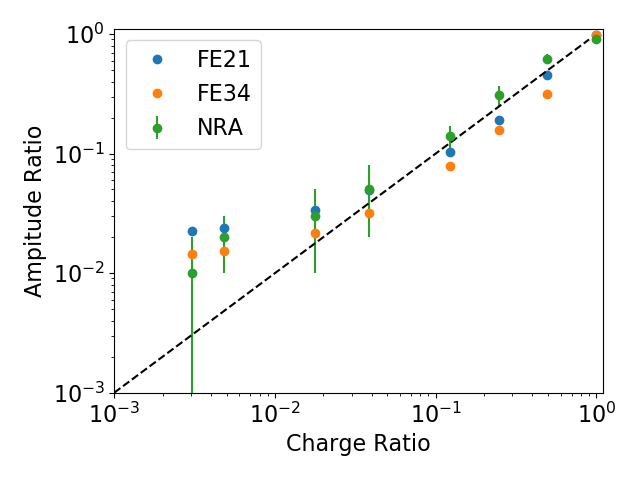}}
\caption{Amplitude ratio of the fitting amplitudes using the FE21 (blue) and FE34 (orange) pinholes, and the NRA (green) analysis, compared with the charge ratio of the excited bunches.}
\label{fig:ratio}
\end{figure}

Using the BBB system, we first applied beam excitation to the first half of the bunches (220). We then gradually reduced the number of excited bunches to 110, 55, 20, 10, 5, 2, and finally to a single bunch.
Figure \ref{fig:ratio} shows the Amplitude ratio defined in Eq. \ref{eq:ratio} obtained with the pinhole data from FE21 and FE34, as well as data from the NRA 7-hole mask, using a double Gaussian model with self-calibration. 
The charge ratio is determined by analyzing the filling pattern using Time-Correlated Single Photon Counting \cite{Torino:2017tpy}.

Using both the pinhole and the NRA mask, we were able to measure the ratio down to $\sim$1\%. To our knowledge, this is the first time an interferometric technique has been used in an accelerator to measure a non-standard beam shape.

\section{Summary}

This proceeding details recent achievements in analyzing data from NRA mask interferometry. We introduce a novel method for calculating beam size that addresses issues with non-uniform illumination by leveraging the closure amplitude technique. We then propose applying this method to wavefront sensing. Finally, we present the first successful measurement of a double Gaussian beam using interferometry in an accelerator, down to a ratio of 1\%. This latter measurement is directly relevant to addressing the core-halo problem in accelerators. All these achievements were obtained using visible light and a non-optimized mask. For future development, we plan to produce new NRA masks specifically tailored to this application to improve results.

%
%
	

%
%



\begin{thebibliography}{9} 

    \bibitem{Nikolic:2024bfx}
        B.~Nikolic, C.~L.~Carilli, N.~Thyagarajan, L.~Torino and U.~Iriso,
        Phys. Rev. Accel. Beams \textbf{27} (2024) no.11, 112802
        doi:10.1103/PhysRevAccelBeams.27.112802
             
    \bibitem{Carilli24}
        C. Carilli, B. Nikolic, L. Torino, U. Iriso, and N. Thyagarajan, ``Deriving the size and shape
        of the ALBA synchrotron light source with optical aperture masking: technical choices'', \emph{Arxiv}, 2024.
        \url{doi:10.48550/arXiv.2406.02114}
        
    \bibitem{Iriso:procibic2024-tup56}
        U. Iriso \emph{et al.},
        in \emph{Proc.  IBIC2024}, Beijing, Sep. 2024, pp. 178-181.
        \url{doi:10.18429/JACoW-IBIC2024-TUP56}

    \bibitem{Mitsuhashi:1998em} 
        T. Mitsuhashi, ``Beam profile and size measurement by SR interferometers,'' in Beam Measurement, May 1999, pp. 399–427. \url{doi:10.1142/9789812818003_0018}
    \bibitem{LT:IBIC14}
        L. Torino and U. Iriso,
        in \emph{Proc. IBIC’15}, Melbourne, Australia, Sep. 2015, pp. 428--432.
        \url{doi:10.18429/JACoW-IBIC2015-TUPB049}  

    \bibitem{1980SPIE..231...18S}
        R. H. Frater, ``Processing of three-dimensional data'', \emph{1980 International Optical Computing
        Conference I , Society of Photo-Optical Instrumentation Engineers (SPIE) Conference Series},
        vol. 231,  p. 18, 1980. \url{doi:10.1117/12.958829}
        
    \bibitem{1987Natur.328..694H}
        C. A. Haniff, C. D. Mackay, D. J. Titterington, D. Sivia, and J. E. Baldwin, ''The first images
        from optical aperture synthesis'', Ann. Sci. Nat. Zool. Biol. Anim., vol. 328, no. 6132, pp. 694–696, Aug. 1987. \url{doi:10.1038/328694a0}

    \bibitem{wikipedia}
        \url{https://en.wikipedia.org/wiki/Gaussian-function}

    \bibitem{Thyagarajan:2025pue}
        N.~Thyagarajan, B.~Nikolic, C.~Carilli, L.~Torino and U.~Iriso,
        ``Two-dimensional Light Beam Shape Characterization using Interferometric Closure Amplitudes,''
        \emph{Journal of the Optical Society of America A,}  Vol. 42, no. 9, pp. 1261-1267 (2025), 
        \url{doi:10.1364/JOSAA.568171}

    \bibitem{PhysRevAccelBeams.19.122801}
        L. Torino and U. Iriso, ``Transverse beam profile reconstruction using synchrotron radiation interferometry,'' \emph{Phys. Rev. Accel. Beams}, vol. 19, no. 12, Dec. 2016. \url{doi:10.1103/physrevaccelbeams.19.122801}
    \bibitem{Safranek:1997mra}
        J.~Safranek,
        Nucl. Instrum. Meth. A \textbf{388} (1997), 27-36
        \url{doi:10.1016/S0168-9002(97)00309-4}
    \bibitem{Carilli:2025dgs}
        C.~L.~Carilli, L.~Torino, B.~Nikolic, N.~Thyagarajan and U.~Iriso,
        \url{arxiv.org/abs/2503.10820}.
    \bibitem{Olmos:2016uwv}
        A.~Olmos, M.~Abbott, U.~Iriso, J.~Moldes, F.~P{\'e}rez, G.~Rehm and I.~Uzun,
        \url{doi:10.18429/JACoW-IBIC2015-TUPB046}

    \bibitem{UI:IBIC22}
    U. Iriso, A. C. Cazorla, I. Mases Solé, A. A. Nosych, and M. Zeus,
    in \emph{Proc. IBIC’22}, Kraków, Poland, Sep. 2022, pp. 421--425.
    \url{doi:10.18429/JACoW-IBIC2022-WEP16} 

    \bibitem{Torino:2017tpy}
       L. Torino, U. Iriso,
      in \emph{Proc. IBIC2016}, Barcelona, Spain, Sep. 2016, paper MOPG59, pp. 202--205,
       \url{doi:10.18429/JACoW-IBIC2016-MOPG59}
    
    \end{thebibliography}
\end{document}